\documentclass[aps,pra,groupedaddress,showpacs,twocolumn]{revtex4}

\usepackage[latin1]{inputenc}   
\usepackage{latexsym}
\usepackage{amssymb}
\usepackage{amsfonts}
\usepackage{bm}
\usepackage{graphicx}
\usepackage{hyperref}

 \providecommand{\dprod}{\!\cdot\!}%
 \providecommand{\wprod}{\!\wedge \!}
 \renewcommand{\pre}[1]{\,^#1\!}
 \providecommand{\sfrac}[2]{\frac{#1}{#2}\,}

\begin{document}


\title{A geometric algebra approach to the hydrogen atom}


\author{José B. Almeida}
\email{bda@fisica.uminho.pt}

\affiliation{Universidade do Minho, Physics Department, Campus de
Gualtar, 4710-057 Braga, Portugal}




\date{\today}

\begin{abstract}
Monogenic functions in the algebra of 5-dimensional spacetime have
been used previously by the author as first principle in different
areas of fundamental physics; the paper recovers that principle
applying it to the hydrogen atom. The equation that results from the
monogenic condition is formally equivalent to Dirac's and so its
solutions resemble closely those found in the literature. The use of
the monogenic condition as point of departure as not only the
advantage of being a unified approach but also provides very strong
links with geometry that are completely lost in the usual approach.
\end{abstract}


\maketitle

\section{Introduction}
I have been advocating in recent papers that the majority of physics
equations can be derived from an appropriately chosen geometry by
exploration of the monogenic condition. Monogenic functions are not
familiar to everybody but they are really the natural extension of
analytic functions when one uses the formalism of geometric algebra
\cite{Doran03, Lasenby99, Hestenes84}; those functions zero the
vector derivative defined on the geometric algebra of the particular
geometry under study.

In \cite{Almeida05:4} I showed how special relativity and the Dirac
equation could be derived from the monogenic condition applied in
the geometric algebra of 5-dimensional spacetime $G_{4,1}$. An
earlier paper \cite{Almeida05:1} proved that the same condition in
the same algebra was sufficient to produce a symmetry group
isomorphic to the standard model gauge group; unfortunately this
paper is incorrect in the formulation of particle dynamics but the
flaw was recently corrected;\cite{Almeida06:2} the same work
introduces electrodynamics and electromagnetism in the monogenic
formalism. Cosmological consequences were drawn from the addition of
an hyperspherical symmetry hypothesis with the consequent choice of
hyperspherical coordinates.\cite{Almeida05:3} Summing up all those
cited papers, I wrote a long book chapter.\cite{Almeida06:3}

The present paper uses the 5D monogenic condition to study electron
orbitals around a positively charged point sized nucleus, a problem
known in quantum mechanics as the "Hydrogen atom problem." The
equation that arises from the monogenic condition applied in
$G_{4,1}$ algebra is formally equivalent to the Dirac equation and
so one expects that its solutions are also formally equivalent to
those obtained in relativistic quantum mechanics. It will be shown,
however, that obtaining the solutions is greatly facilitated by the
use of geometric algebra formalism and also that the resulting
formulas are much more compact than standard ones and lend
themselves to an easier geometrical interpretation. The derivations
make use of some methods and strategies drawn from Refs.\
\cite{Doran03, Lasenby99} but depart from those works in many
important aspects.

\section{Some geometric algebra \label{somealg}}
Geometric algebra is not usually taught in university courses and
its presence in the literature is scarce; good reference works are
\cite{Doran03, Hestenes84, Lasenby99}. We will concentrate on the
algebra of 5-dimensional spacetime because this will be our main
working space; this algebra incorporates as subalgebras those of the
usual 3-dimensional Euclidean space, Euclidean 4-space and Minkowski
spacetime. We begin with the simpler 5-D flat space and progress to
a 5-D spacetime of general curvature (see Appendix \ref{galgebra}
for more details.)

The geometric algebra ${G}_{4,1}$ of the hyperbolic 5-dimensional
space we consider is generated by the coordinate frame of
orthonormal basis vectors $\sigma_\alpha $ such that
\begin{eqnarray}
\label{eq:basis}
    && (\sigma_0)^2  = -1, \nonumber \\
    && (\sigma_i)^2 =1, \\
    && \sigma_\alpha \dprod \sigma_\beta  =0, \quad \alpha \neq \beta. \nonumber
    \nonumber
\end{eqnarray}
Note that the English characters $i, j, k$ range from 1 to 4 while
the Greek characters $\alpha, \beta, \gamma$ range from 0 to 4. See
Appendix \ref{indices} for the complete notation convention used.

Any two basis vectors can be multiplied, producing the new entity
called a bivector. This bivector is the geometric product or, quite
simply, the product, and it is distributive. Similarly to the
product of two basis vectors, the product of three different basis
vectors produces a trivector and so forth up to the fivevector,
because five is the dimension of space.

We will simplify the notation for basis vector products using
multiple indices, i.e.\ $\sigma_\alpha \sigma_\beta \equiv
\sigma_{\alpha\beta}.$ The algebra is 32-dimensional and is spanned
by the basis
\begin{itemize}
\item 1 scalar, { $1$},
\item 5 vectors, { $\sigma_\alpha$},
\item 10 bivectors (area), { $\sigma_{\alpha\beta}$},
\item 10 trivectors (volume), { $\sigma_{\alpha\beta\gamma}$},
\item 5 tetravectors (4-volume), { $\mathrm{i} \sigma_\alpha $},
\item 1 pseudoscalar (5-volume), { $\mathrm{i} \equiv
\sigma_{01234}$}.
\end{itemize}
Several elements of this basis square to unity:
\begin{equation}
\label{eq:positive}
    (\sigma_i)^2 =  (\sigma_{0i})^2=
    (\sigma_{0i j})^2 =(\mathrm{i}\sigma_0)^2 =1.
\end{equation}
The remaining basis elements square to $-1$:
\begin{equation}
    \label{eq:negative}
    (\sigma_0)^2 = (\sigma_{ij})^2 = (\sigma_{ijk})^2 =
    (\mathrm{i}\sigma_i)^2 = \mathrm{i}^2=-1.
\end{equation}
Note that the pseudoscalar $\mathrm{i}$ commutes with all the other
basis elements while being a square root of $-1$; this makes it a
very special element which can play the role of the scalar imaginary
in complex algebra.

In 5-dimensional spacetime of general curvature, spanned by 5
coordinate frame vectors $g_\alpha$, the indices follow the
conventions set forth in Appendix \ref{indices}. We will also assume
this spacetime to be a metric space whose metric tensor is given by
\begin{equation}
\label{eq:metrictens}
    g_{\alpha \beta} = g_\alpha \dprod g_\beta;
\end{equation}
the double index is used with $g$ to denote the inner product of
frame vectors and not their geometric product. The space signature
is $(-++++)$, which amounts to saying that $g_{00} < 0$ and $g_{ii}
>0$. If the coordinate frame vectors can be expressed as a linear
combination of the orthonormed ones, we have
\begin{equation}
    \label{eq:indexframemain}
    g_\alpha = {n^\beta}_\alpha \sigma_\beta,
\end{equation}
where ${n^\beta}_\alpha$ is called the \emph{refractive index
tensor} or simply the \emph{refractive index}; its 25 elements can
vary from point to point as a function of the
coordinates.\cite{Almeida04:4} In this work we will not consider
spaces of general curvature but only those verifying condition
(\ref{eq:indexframemain}); in those spaces we define the vector and
covariant derivatives (see appendix \ref{derivatives}).

\section{Dirac's equation}

There is a class of functions of great importance, called
\emph{monogenic functions,}\cite{Doran03} characterized by having
null vector derivative; a function $\Psi$ is monogenic in flat space
if and only if
\begin{equation}
    \label{eq:monogenic}
    \nabla \Psi = 0.
\end{equation}
A monogenic function is not usually a scalar and has by necessity
null Laplacian, as can be seen by dotting Eq.\ (\ref{eq:monogenic})
with $\nabla$ on the left; it has solutions of the type
\begin{equation}
    \label{eq:psidef}
    \Psi = \Psi_0 \mathrm{e}^{\mathrm{i} (p_\alpha x^\alpha +
    \delta)}.
\end{equation}
The Dirac equation can be derived from the monogenic condition, as
shown in \cite{Almeida05:4} and briefly remembered here. For this
effect it will be convenient to expand the monogenic condition
(\ref{eq:monogenic}) as $(\pre{\mu}\nabla + \sigma^4
\partial_4) \Psi = 0$. If this is applied to the solution
(\ref{eq:psidef}) and the derivative with respect to $x^4$ is
evaluated we get
\begin{equation}
\label{eq:splitmono}
    (\pre{\mu}\nabla + \sigma^4 \mathrm{i} p_4) \Psi = 0.
\end{equation}
Let us now multiply both sides of the equation on the left by
$\sigma^4$ and note that bivector $\sigma^{40}$ squares to the
identity while the 3 bivectors $\sigma^{4m}$ square to minus
identity; we rename these bivectors as $\gamma$-bivectors in the
form $\gamma^\mu = \sigma^{4\mu}$. Rewriting the equation in this
form we get
\begin{equation}
\label{eq:Diracgamma}
    (\gamma^\mu \partial_\mu + \mathrm{i} p_4) \Psi = 0.
\end{equation}
Invoking the isomorphism between $G_{4,1}$ and the complex algebra
of $4 \ast 4$ matrices, the only thing this equation needs to be
recognized as Dirac's is the replacement of $p_4$ by the particle's
mass $m$; simultaneously we assign the energy $E$ to $p_0$ and 3D
momentum $\mathbf{p}$ to $\sigma^m p_m$.\cite{Almeida05:4,
Almeida06:2} Alternatively we can multiply both sides of Eq.\
(\ref{eq:splitmono}) by $\sigma^0$ on the left to obtain the
$\alpha$, $\beta$ form of the Dirac equation, through the assignment
$\alpha^m \equiv \sigma^{m0}$, $\beta \equiv \sigma^{40}$.

Applying the monogenic condition to Eq.\ (\ref{eq:psidef}) we see
that the following equation must be verified
\begin{equation}
    \Psi_0 (\sigma^\alpha p_\alpha) = 0.
\end{equation}
It has been shown\cite{Almeida05:4} that $p = \sigma^\alpha
p_\alpha$ is a null vector and the wavefunction in Eq.\
(\ref{eq:psidef}) can be given a different form, taking in
consideration the previous assignments
\begin{equation}
    \label{eq:fermion}
    \Psi = A(\pm \sigma_4 m + \mathbf{p} + \sigma_0 E)
     \mathrm{e}^{\mathrm{i} ( E t + \mathbf{p}\cdot \mathbf{x} \pm m \tau + \delta)};
\end{equation}
where $A$ is the amplitude and $\mathbf{x} = \sigma_m x^m$ is the
3-dimensional position. Although $\Psi$ does not look like a column
Dirac spinor, it has the same number of components and can be
written in that form, if desired.

In Eq.\ (\ref{eq:splitmono}) we used $\mathrm{i}$ as imaginary in
the exponent but the monogenic condition would be equally verified
if we had chosen any other algebra element whose square was minus
unity; in \cite{Almeida05:1} the various possible such elements are
analysed and discussed; in the present work we will not explore such
possibility. Applying the vector derivative to this solution we have
\begin{equation}
\label{eq:gennabla}
    \nabla \Psi = (E \sigma^0 \pm m \sigma^4)\mathrm{i} \Psi +
    \bm{\nabla} \Psi = 0.
\end{equation}

\section{Electromagnetic potentials}

When working in curved spaces the monogenic condition is naturally
modified, replacing the vector derivative $\nabla$ with the
covariant derivative $\mathrm{D}$. A generalized monogenic function
is then a function that verifies the equation
\begin{equation}
\label{eq:genmonogenic}
    \mathrm{D} \Psi = 0.
\end{equation}

Remembering that an electron has minus unit charge in our units'
system, for an electron with rest mass $m$ in the presence of an
electromagnetic potential $A$ we have to consider the reciprocal
frame (see \cite{Almeida06:2} and \cite{Almeida05:3})
\begin{equation}
    g^\mu = \sigma^\mu, ~~~~ g^4 = \frac{-A_\mu}{m}\, \sigma^\mu + \sigma^4,
\end{equation}
which corresponds to the refractive index frame
\begin{equation}
    g_0 = \sigma_\mu + \frac{A_\mu}{m}\, \sigma_4,~~~~g_4 = \sigma_4;
\end{equation}
it is easily verified that $g^\alpha g_\beta =
{\delta^\alpha}_\beta$, as required by the definition of reciprocal
frame. The covariant derivative is then
\begin{equation}
    \mathrm{D} = \sigma^0 \partial_0 + \bm{\nabla}
    + \left(-\frac{A_\mu}{m}\, \sigma^\mu + \sigma^4\right) \partial_4.
\end{equation}
We expect solutions of Eq.\ (\ref{eq:genmonogenic}) that are
harmonic in $t$ and $\tau$, which we write as $\Psi = \psi(x^m)
\exp[\mathrm{i}( E t - m \tau)]$, selecting the signs in the
exponent for a forward propagating wave. The monogenic condition
becomes
\begin{equation}
    \left[\sigma^0  (E -A_0) -\bm{\nabla}\mathrm{i}
    -\sigma^m A_m  - \sigma^4 m\right] \Psi = 0.
\end{equation}

For the following derivations we will follow closely the procedure
explained in detail in \citet{Doran03}, making the necessary changes
to conform to our monogenic formalism. In the case of a central
field we have to make $A_m = 0$ and $A_0 = \phi(r)$, with $r$ the
radial coordinate and the previous equation becomes
\begin{equation}
\label{eq:central}
    \left[\sigma^0  (E -\phi) -  \bm{\nabla}\mathrm{i}
      - \sigma^4 m\right] \Psi = 0.
\end{equation}
Under those conditions we write the previous equation as
\begin{eqnarray}
\label{eq:almeidian}
    &&\bm{\nabla} \Psi = \mathcal{D}\Psi, \nonumber \\
    && \mathcal{D} = - (E \sigma^0  - \phi \sigma^0 -
    {m} \sigma^4)\mathrm{i} .
\end{eqnarray}
The wavefunction $\Psi$ is multiplied by a vector on both sides of
Eq.\ (\ref{eq:almeidian}) but the vector on the lhs has only 3D
components while the vector on the rhs has only $\sigma^0$ and
$\sigma^4$ components; this arrangement is particularly useful to
study commutativity of operators, as we shall see. Suppose we have
an operator $\mathcal{O}$ that  commutes  with both $\bm{\nabla}$
and $\mathcal{D}$ and suppose also that $\Psi$ is in an eigenstate
of $\mathcal{O}$, that is $\mathcal{O}\Psi = o \Psi$; Eq.\
(\ref{eq:almeidian}) is then automatically verified. Our task is
then to find the operators that verify the commutation conditions
and solve for their eigenstates; this procedure is very similar to
what is done in quantum mechanics for the hydrogen atom.

In Appendix \ref{commute} we define the total angular momentum $J_3$
whose eigenstates define the angular solutions; another associated
operator is
\begin{equation}
    \mathcal{K} = (1 - \mathbf{x}\wprod
    \bm{\nabla})\sigma^{40}.
\end{equation}
$\mathcal{K}$ anti-commutes with $\mathcal{D}$ because the bivectors
present in $\mathbf{x}\wprod \bm{\nabla}$ belong to 3D space and
$\sigma^{40}$ anti-commutes with both $\sigma^0$ and $\sigma^4$. The
fact that it does not commute with $\mathcal{D}$ is not a problem as
long as $\mathcal{K}$ has symmetric eigenvalues; it commutes with
$\bm{\nabla}$ as shown in Appendix \ref{commute}. So, we can assume
$\Psi$ to be in an eigenstate of $\mathcal{K}$, that is
\begin{equation}
    \mathcal{K} \Psi = k \Psi.
\end{equation}

With central potentials we will use spherical coordinates; this
implies that the frame vectors undergo rotations and their
derivatives must be considered as follows
\begin{equation}
\begin{array}{ll}
    \partial_\theta \sigma_r = \sigma_\theta,~~
    & \partial_\varphi \sigma_r = \sin \theta \sigma_\varphi,  \\
    \partial_\theta \sigma_\theta =
    -\sigma_r,~~ & \partial_\varphi \sigma_\theta = \cos \theta \sigma_\varphi, \\
    \partial_\theta \sigma_\varphi = 0,~~ &
    \partial_\varphi \sigma_\varphi = -\sin \theta\, \sigma_r - \cos \theta\, \sigma_\theta.
\end{array}
\end{equation}
The 3D part of the vector derivative becomes
\begin{equation}
    \bm{\nabla} = \sigma^r \partial_r + \sfrac{1}{r}
    (\sigma^\theta \partial_\theta + \csc \theta \,\sigma^\varphi
    \partial_\varphi).
\end{equation}

\section{Angular solutions}
In order to solve Eq.\ (\ref{eq:central})  we make the ansatz $\Psi
= R Y^s_t$, where $Y^s_t$ contains all the angular dependence and
$R$ is a function of $r$, $t$ and $\tau$ only. We note also that
$\mathbf{x}\wprod \bm{\nabla} R = 0$; in terms of this operator we
can write
\begin{equation}
\label{eq:1minusk}
    \mathbf{x}\wprod \bm{\nabla} \Psi = (1- \sigma^{40} k) \Psi
    = R\, \mathbf{x}\wprod \bm{\nabla} Y^s_t.
\end{equation}

The analysis starts with 3D monogenic functions, or spherical
harmonics, defined by $\bm{\nabla} \psi^s_t=0$; these functions are
of the type $\psi^s_t = r^{t} Y^s_t(\theta, \varphi)$ and the vector
derivative is
\begin{equation}
    \bm{\nabla} \psi^s_t = t \sigma^r r^{t-1} Y^s_t +  r^{t} \bm{\nabla} Y^s_t.
\end{equation}
Since the first member must be null
\begin{equation}
\label{eq:angular}
    \bm{\nabla} Y^s_t = -\frac{t}{r}\, \sigma^r Y^s_t.
\end{equation}
In terms of operator $\mathbf{x}\wprod \bm{\nabla}$ the equation is
\begin{equation}
\label{eq:yst}
    -\mathbf{x}\wprod \bm{\nabla} Y^s_t = t Y^s_t;
\end{equation}
comparing with Eq.\ (\ref{eq:1minusk}) we see that $k = t+1$.
Equation (\ref{eq:yst}) is an eigenvalue equation satisfied by
spherical harmonics with the general formula\cite{Doran03}
\begin{eqnarray}
    Y^s_t &=& \left[(s+t+1)P^s_t(\cos \theta) \right. \nonumber \\
    && \left. - P^{s+1}_t(\cos \theta) \sigma^{r \theta}\right]
    \mathrm{e}^{s \varphi \sigma^{1 2}},
\end{eqnarray}
where $P^s_t(x)$ are Legendre polynomials, $t \geq 0$ and $-1-t \leq
s \leq t$; solutions for $t \leq -2$ can be found through the
relation
\begin{equation}
    -\mathbf{x}\wprod \bm{\nabla} (\sigma^r \psi \sigma^3) = - (t+2)
    \sigma^r \psi \sigma^3.
\end{equation}
Because $k = t+1$, the allowed values for $k$ are given by $|k|>0$;
this ensures that the eigenvalues of $\mathcal{K}$ operator are
indeed symmetric, as we required in the previous section.

A particularly simple formula for 3D monogenic functions is obtained
when the two quantum numbers are identical, in which case we obtain
\begin{equation}
    \psi^t_t = (r \sin \theta)^t \mathrm{e}^{-t \varphi \sigma^{12}}.
\end{equation}
Since Eq.\ (\ref{eq:angular}) does not depend on $s$, we can always
use this particular case for the discussion of radial solutions.

\section{Radial solutions}
For an hydrogen-like atom we make $\phi = Z\alpha/r$, with $Z$ the
positive charge of the nucleus and $\alpha$ the fine structure
constant. We will multiply Eq.\ (\ref{eq:almeidian}) on the left by
$\mathbf{x}$ noting that $\mathbf{x}\bm{\nabla} = r\partial_r +
\mathbf{x}\wprod \bm{\nabla}$; considering Eq.\ (\ref{eq:1minusk})
\begin{equation}
    r \partial_r \Psi + (1- k \sigma^{40}) \Psi
     =   r  \left(E  \sigma^{0r}
    - \sfrac{Z \alpha}{r} \sigma^{0r} -
    {m} \sigma^{4r} \right)\mathrm{i}\Psi.
\end{equation}
Introducing the function $\Upsilon = r \Psi$, this can be rearranged
isolating $\partial_r \Upsilon$
\begin{equation}
    \partial_r \Upsilon =  (E  \sigma^{0r} +
    {m} \sigma^{4r} )\mathrm{i} \Upsilon + \sfrac{1}{r}(k \sigma^{40}
     +Z \alpha \sigma^{r0}\mathrm{i})\Upsilon.
\end{equation}
It is useful to define the two multivectors
\begin{eqnarray}
    F &=& - (E  \sigma^{0r} -
    {m} \sigma^{4r} )\mathrm{i}, \nonumber\\
    G &=& -(k \sigma^{40}
     +Z \alpha \sigma^{r0}\mathrm{i});
\end{eqnarray}
so that the monogenic condition becomes
\begin{equation}
    \partial_r \Upsilon + \left(F + \sfrac{G}{r} \right)\Upsilon.
\end{equation}

The $F$ and $G$ operators satisfy
\begin{eqnarray}
    F^2 &=& m^2 - E^2 = f^2, \nonumber\\
    G^2 &=& k^2 - (Z\alpha)^2 = \nu^2.
\end{eqnarray}
They also observe the anticommutation relation
\begin{equation}
    FG + GF = -2Z\alpha E.
\end{equation}

We will now make a change of variable to allow the separation of
large and small $r$ behaviour; this is $y = fr$ with which we write
\begin{equation}
    \Upsilon = \Phi \mathrm{e}^{-y}.
\end{equation}
The new function $\Phi$ satisfies
\begin{equation}
    \partial_r \Phi + \sfrac{G}{y}\Phi + \left(\sfrac{F}{f} - 1 \right) \Phi = 0.
\end{equation}
We can certainly express $\Phi$ as a power series; moreover this
power series must not be infinite, otherwise it would not fall to
zero at large $r$. Calling $C_l$ to the series coefficients
\begin{equation}
    \Phi = y^s \sum_{l=0}^{n} C_l y^l.
\end{equation}
The coefficients verify the recursion relation
\begin{equation}
    (l + s + G)C_l = - \left(\sfrac{F}{f} -1 \right) C_{l-1}.
\end{equation}
For $l = 0$ it is
\begin{equation}
    (s + G)C_0 = 0.
\end{equation}
Multiplying on the left by $(s-G)$ we can see that $s^2 = G^2 =
\nu^2$ and we set $s = \nu$ to avoid a central singularity.

Since the series terminates at $l=n$, the coefficient $C_{n+1}$ must
be null but it must still verify the recursion relation
\begin{equation}
    \left(\sfrac{F}{f} - 1 \right) C_n = 0,
\end{equation}
and so $FC_n = f C_n$. Multiplying both sides of the recursion
relation on the left by (F/f + 1) and replacing $s$ by $\nu$
\begin{eqnarray}
    &&\left(\sfrac{F}{f} + 1 \right)(n + \nu + G)C_n = \nonumber \\
    &&=  -   \left(\sfrac{F}{f} + 1 \right) \left(\sfrac{F}{f} - 1
    \right)C_{n-1} = 0,
\end{eqnarray}
Combining the two equations we get
\begin{equation}
    \left[2 (n + \nu) + G + \sfrac{F}{f} G \right] C_n = 0.
\end{equation}
This can in turn be manipulated to give
\begin{equation}
    \left[2 (n + \nu) + \sfrac{1}{f} (GF + FG) \right] = 0;
\end{equation}
and finally
\begin{equation}
    n + \nu - \sfrac{Z \alpha E}{f} = 0.
\end{equation}
This is the energy quantization equation, which can be arranged into
the usual form by first manipulating to
\begin{equation}
    \sfrac{E}{\sqrt{m^2 - E^2}} = \sfrac{n + \nu}{Z\alpha},
\end{equation}
and then rearranging to
\begin{equation}
    E^2 = m^2 \left[1- \sfrac{(Z \alpha)^2}{n^2 + 2n\nu + k^2}
    \right].
\end{equation}

\section{Ground state of the hydrogen atom}

It is useful to analyse the ground state in order to find the
character of the series coefficients. Since this is a spherically
symmetric solution we ignore the $\theta$ and $\varphi$ dependence.
At large $r$ we can neglect the angular dependence of $\sigma^r$ and
a second order equation can be written for $\psi(x^m)$ as
\begin{equation}
    \partial_{rr}\psi \approx \left(E \sigma^0  -    m \sigma^4\right)
    \sigma^r \partial_r \psi \mathrm{i} = (m^2 - E^2) \psi.
\end{equation}
Since we are looking for bound states it must be $E < m$ and $\psi$
goes with $\exp(-f r)$, with $f = \sqrt{m^2-E^2}$; the large $r$
solution is
\begin{equation}
    \psi = \psi_l \mathrm{e}^{-f r}.
\end{equation}
Inserting into Eq.\ (\ref{eq:genmonogenic})
\begin{equation}
    f \sigma^r \psi_l  = (E\sigma^0 - m \sigma^4)\mathrm{i}\psi_l.
\end{equation}
The equation is solved if $\psi_l$ contains a factor $(E \sigma^{0}
-m \sigma^{4}) + f \sigma^r\mathrm{i} $.

For small $r$ Eq.\ (\ref{eq:genmonogenic}) becomes
\begin{equation}
\label{eq:smallr}
    \bm{\nabla} \psi \approx -\sfrac{Z \alpha}{r} \sigma^0 \psi \mathrm{i}.
\end{equation}
Because we are assuming $\psi$ to be a radial function we try a
solution of the type
\begin{equation}
    \psi = a \sigma^r + b \sigma^0,
\end{equation}
where $a$ and $b$ are scalar functions of $r$. We note that $\nabla
\sigma^r = 2/r$ and insert in Eq.\ (\ref{eq:smallr}) to get
\begin{eqnarray}
    a' + \sfrac{2a}{r} + b' \sigma^{r0} &= - \sfrac{Z \alpha}{r}
    \sigma^0 (a \sigma^r + b \sigma^0)\mathrm{i} \nonumber \\
    &= (a \sigma^{r0} + b)\sfrac{Z \alpha}{r}\mathrm{i}.
\end{eqnarray}
This implies the simultaneous equations
\begin{equation}
\label{eq:a_and_b}
    \left\{\begin{array}{l}
    \displaystyle
    a' + \sfrac{2a}{r} = \sfrac{Z \alpha b}{r} \mathrm{i}\\  \\
    \displaystyle
    b' = \sfrac{Z \alpha a}{r} \mathrm{i}
    \end{array} \right.
\end{equation}
From the first equation in this set we take the derivative of $b$ as
\begin{equation}
    b' = -(3 a' + r a^{''})\sfrac{\mathrm{i}}{Z \alpha}.
\end{equation}
We can now combine the two equations for $b'$ into the single
differential equation
\begin{equation}
    \sfrac{Z \alpha a}{r} = - \sfrac{3 a' + r a^{''}}{Z \alpha},
\end{equation}
which we solve as
\begin{equation}
    a = r^\eta,
\end{equation}
with
\begin{equation}
    \eta = -1 \pm \sqrt{1 - (Z \alpha)^2};
\end{equation}
we will later argue that only the plus sign is physically
meaningful. Inserting $a$ into the second Eq.\ (\ref{eq:a_and_b})
and solving we get for $b$
\begin{equation}
    b= \sfrac{Z \alpha}{\eta} r^\eta.
\end{equation}

Actually it is convenient to multiply $a$ and $b$ by $\mathrm{i}$.
Summarizing the results for small and large $r$, respectively,
\begin{equation}
\label{eq:summary1}
    \psi = \left\{\begin{array}{ll}
        \displaystyle
    \left(\sfrac{Z \alpha}{-\eta}
    \sigma^0 + \sigma^r \mathrm{i} \right) r^\eta, & r \text{
    small},\\ & \\
    \displaystyle
    \left(\sfrac{E \sigma^0 - m \sigma^4}{f} + \sigma^r \mathrm{i} \right)
    \mathrm{e}^{-f r}, & r \text{ large}.
    \end{array} \right .
\end{equation}
In order to make the two factors in brackets compatible we must
include a factor $(1-\sigma^{40})$ to get
\begin{equation}
    \psi = \left\{\begin{array}{ll}
        \displaystyle
    \left(\sfrac{Z \alpha}{-\eta}
    \sigma^0 + \sigma^r \mathrm{i} \right)(1-\sigma^{40}) r^\eta, & \\ & \\ \displaystyle
    \left(\sfrac{E \sigma^0 - m \sigma^4}{f} + \sigma^r \mathrm{i} \right)(1-\sigma^{40})
    \mathrm{e}^{-f r}. &
    \end{array} \right.
\end{equation}
\begin{equation}
    \psi = \left\{\begin{array}{ll}
        \displaystyle
    \left[\sfrac{Z \alpha}{-\eta}
    (\sigma^0- \sigma^4) +(1-\sigma^{40}) \sigma^r \mathrm{i} \right] r^\eta, &\\ & \\ \displaystyle
    \left[\sfrac{E  + m }{f}(\sigma^0 - \sigma^4) + (1-\sigma^{40})\sigma^r \mathrm{i} \right]
    \mathrm{e}^{-f r}, &
    \end{array} \right.
\end{equation}

The complete radial solution for the ground state is
\begin{equation}
    \psi = \left[\sfrac{Z \alpha}{-\eta}
    (\sigma^0- \sigma^4) +(1-\sigma^{40}) \sigma^r \mathrm{i} \right] r^\eta\mathrm{e}^{-f r}.
\end{equation}
under the condition
\begin{equation}
    \sfrac{E + m}{f} = - \sfrac{Z \alpha}{\eta} = \sfrac{Z \alpha}{1
    \mp \sqrt{1 - (Z \alpha)^2}}.
\end{equation}
The left hand side can be written as
\begin{equation}
    \sfrac{E + m}{f} = \sfrac{f/m}{1 - E/m} = \sfrac{f/m}{1 \mp
    \sqrt{1 - (f/m)^2}}.
\end{equation}
In order to have a positive energy it must be $f = m Z \alpha$ and
\begin{equation}
    E = m \sqrt{1 - (Z \alpha)^2}\,.
\end{equation}
This is a particular case of the general energy quantization formula
derived above, for $n=0$ and $k = 1$.

\section{Conclusion}
The relativistic solutions for the hydrogen atom can be derived from
the monogenic condition applied to functions in the algebra of 5D
spacetime. When the space is bent by the consideration of a central
potential, the equation that follow from the monogenic condition is
formally equivalent to Dirac's and so one could expect equivalent
solutions from the onset. The paper revisits the solutions of that
equation applying the formalism consistent with the monogenic
condition.

There is nothing fundamentally new in the paper, in the sense that
the energy levels that one obtains with the monogenic formalism are
the same that can be found in the literature. However, the monogenic
condition had previously been applied in other areas of physics
showing high unifying potential; this work is just one further step
in the path of unification. Besides that, the author believes the
monogenic formalism to be easier to apprehend than the more usual
matrix formalism; this is a question of taste, though.

\begin{appendix}
\section{Indexing conventions \label{indices}}
\begin{figure}[htb]
\vspace{11pt} \centerline{\includegraphics[scale=1]{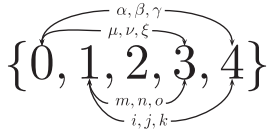}}

\caption{\label{f:indices} Indices in the range $\{0,4\}$ will be
denoted with Greek letters $\alpha, \beta, \gamma.$ Indices in the
range $\{0,3\}$ will also receive Greek letters but chosen from
$\mu, \nu, \xi.$ For indices in the range $\{1,4\}$ we will use
Latin letters $i, j, k$ and finally for indices in the range
$\{1,3\}$ we will use also Latin letters chosen from $m, n, o.$ }
\end{figure}

In this section we establish the indexing conventions used in the
paper. We deal with 5-dimensional space but we are also interested
in two of its 4-dimensional subspaces and one 3-dimensional
subspace; ideally our choice of indices should clearly identify
their ranges in order to avoid the need to specify the latter in
every equation. The diagram in Fig.\ \ref{f:indices} shows the index
naming convention used in this paper;
Einstein's summation convention will be adopted as well as the
compact notation for partial derivatives $\partial_\alpha =
\partial/\partial x^\alpha.$

\section{Non-dimensional units \label{units}}
The interpretation of $t$ and $\tau$ as time coordinates implies the
use of a scale parameter which is naturally chosen as the vacuum
speed of light $c$. We don't need to include this constant in our
equations because we can always recover time intervals, if needed,
introducing the speed of light at a later stage. We can even go a
step further and eliminate all units from our equations so that they
become pure number equations; in this way we will avoid cumbersome
constants whenever coordinates have to appear as arguments of
exponentials or trigonometric functions. We note that, at least for
the macroscopic world, physical units can all be reduced to four
fundamental ones; we can, for instance, choose length, time, mass
and electric charge as fundamental, as we could just as well have
chosen others. Measurements are then made by comparison with
standards; of course we need four standards, one for each
fundamental unit. But now note that there are four fundamental
constants: Planck constant $(\hbar)$, gravitational constant $(G)$,
speed of light in vacuum $(c)$ and proton electric charge $(e)$,
with which we can build four standards for the fundamental units.
\begin{table}[htb]
\caption{\label{t:standards}Standards for non-dimensional units'
system}
\begin{center}
\begin{tabular}{c|c|c|c}
Length & Time & Mass & Charge \\
\hline & & & \\

$\displaystyle \sqrt{\frac{G \hbar}{c^3}} $ & $\displaystyle
\sqrt{\frac{G \hbar}{c^5}} $  & $\displaystyle \sqrt{\frac{ \hbar c
}{G}} $  & $e$
\end{tabular}
\end{center}
\end{table}
Table \ref{t:standards} lists the standards of this units' system,
frequently called Planck units, which the authors prefer to
designate by non-dimensional units. In this system all the
fundamental constants, $\hbar$, $G$, $c$, $e$, become unity, a
particle's Compton frequency, defined by $\nu = mc^2/\hbar$, becomes
equal to the particle's mass and the frequent term ${GM}/({c^2 r})$
is simplified to ${M}/{r}$. We can, in fact, take all measures to be
non-dimensional, since the standards are defined with recourse to
universal constants; this will be our posture. Geometry and physics
become relations between pure numbers, vectors, bivectors, etc. and
the geometric concept of distance is needed only for graphical
representation.

\section{Some complements of geometric algebra \label{galgebra}}
In this section we expand the concepts given in Sec.\ \ref{somealg},
introducing some useful relations and definitions. Starting with the
basis elements that square to unity Eq.\ (\ref{eq:positive}),
repeated here,
\begin{equation}
    (\sigma_i)^2 =  (\sigma_{0i})^2=
    (\sigma_{0i j})^2 =(\mathrm{i}\sigma_0)^2 =1,
\end{equation}
it is easy to verify any of the above equations; suppose we want to
check that $(\sigma_{0i j})^2 = 1$. Start by expanding the square
and remove the compact notation $(\sigma_{0i j})^2 = \sigma_0
\sigma_i \sigma_j \sigma_0 \sigma_i \sigma_j$, then swap the last
$\sigma_j$ twice to bring it next to its homonymous; each swap
changes the sign, so an even number of swaps preserves the sign:
$(\sigma_{0i j})^2 = \sigma_0 \sigma_i (\sigma_j)^2 \sigma_0
\sigma_i$. From the third equation (\ref{eq:basis}) we know that the
squared vector is unity and we get successively $(\sigma_{0i j})^2 =
\sigma_0 \sigma_i \sigma_0 \sigma_i = - (\sigma_0)^2 (\sigma_i)^2 =
- (\sigma_0)^2 $; using the first equation (\ref{eq:basis}) we get
finally $(\sigma_{0i j})^2 = 1$ as desired.

The remaining basis elements square to $-1$ as can be verified in a
similar manner, Eq.\ (\ref{eq:negative}):
\begin{equation}
    (\sigma_0)^2 = (\sigma_{ij})^2 = (\sigma_{ijk})^2 =
    (\mathrm{i}\sigma_i)^2 = \mathrm{i}^2=-1.
\end{equation}
Note that the pseudoscalar $\mathrm{i}$ commutes with all the other
basis elements while being a square root of $-1$; this makes it a
very special element which can play the role of the scalar imaginary
in complex algebra.

We can now address the geometric product of any two vectors $a =
a^\alpha \sigma_\alpha$ and $b = b^\beta \sigma_\beta$ making use of
the distributive property
\begin{equation}
    ab = \left(-a^0 b^0 + \sum_i a^i b^i \right) + \sum_{\alpha \neq \beta}
    a^\alpha b^\beta \sigma_{\alpha \beta};
\end{equation}
and we notice it can be decomposed into a symmetric part, a scalar
called the inner or interior product, and an anti-symmetric part, a
bivector called the outer or exterior product.
\begin{equation}
    ab = a \dprod b + a \wprod b,~~~~ ba = a \dprod b - a \wprod b.
\end{equation}
Reversing the definition one can write inner and outer products as
\begin{equation}
    a \dprod b = \frac{1}{2}\, (ab + ba),~~~~ a \wprod b = \frac{1}{2}\, (ab -
    ba).
\end{equation}
The inner product is the same as the usual ''dot product,'' the only
difference being in the negative sign of the $a_0 b_0$ term; this is
to be expected and is similar to what one finds in special
relativity. The outer product represents an oriented area; in
Euclidean 3-space it can be linked to the "cross product" by the
relation $\mathrm{cross}(\mathbf{a},\mathbf{b}) = - \sigma_{123}
\mathbf{a} \wprod \mathbf{b}$; here we introduced bold characters
for 3-dimensional vectors and avoided defining a symbol for the
cross product because we will not use it again. We also used the
convention that interior and exterior products take precedence over
geometric product in an expression.

When a vector is operated with a multivector the inner product
reduces the grade of each element by one unit and the outer product
increases the grade by one. We will generalize the definition of
inner and outer products below; under this generalized definition
the inner product between a vector and a scalar produces a vector.
Given a multivector $a$ we refer to its grade-$r$ part by writing
$<\!a\!>_r$; the scalar or grade zero part is simply designated as
$<\!a\!>$. By operating a vector with itself we obtain a scalar
equal to the square of the vector's length
\begin{equation}
    a^2 = aa = a \dprod a + a \wprod a = a \dprod a.
\end{equation}
The definitions of inner and outer products can be extended to
general multivectors
\begin{eqnarray}
    a \dprod b &=& \sum_{\alpha,\beta} \left<<\!a\!>_\alpha \;
    <\!b\!>_\beta \right>_{|\alpha-\beta|},\\
    a \wprod b &=& \sum_{\alpha,\beta} \left<<\!a\!>_\alpha \;
    <\!b\!>_\beta \right>_{\alpha+\beta}.
\end{eqnarray}
Two other useful products are the scalar product, denoted as
$<\!ab\!>$ and commutator product, defined by
\begin{equation}
    a \times b  = \sfrac{1}{2}(ab - ba).
\end{equation}
In mixed product expressions we will use the convention that inner
and outer products take precedence over geometric products as said
above.

We will encounter exponentials with multivector exponents; two
particular cases of exponentiation are specially important. If $u$
is such that $u^2 = -1$ and $\theta$ is a scalar
\begin{eqnarray}
   \mathrm{e}^{u \theta} &=& 1 + u \theta -\frac{\theta^2}{2!} - u
    \frac{\theta^3}{3!} + \frac{\theta^4}{4!} + \ldots  \nonumber \\
    &=& 1 - \frac{\theta^2}{2!} +\frac{\theta^4}{4!}- \ldots \{=
    \cos \theta \} \nonumber \\
    && + u \theta - u \frac{\theta^3}{3!} + \ldots \{= u \sin
    \theta\} \\
    &=&  \cos \theta + u \sin \theta. \nonumber
\end{eqnarray}
Conversely if $h$ is such that $h^2 =1$
\begin{eqnarray}
    \mathrm{e}^{h \theta} &=& 1 + h \theta +\frac{\theta^2}{2!} + h
    \frac{\theta^3}{3!} + \frac{\theta^4}{4!} + \ldots  \nonumber \\
    &=& 1 + \frac{\theta^2}{2!} +\frac{\theta^4}{4!}+ \ldots \{=
    \cosh \theta \} \nonumber \\
    && + h \theta + h \frac{\theta^3}{3!} + \ldots \{= h \sinh
    \theta\}  \\
    &=&  \cosh \theta + h \sinh \theta. \nonumber
\end{eqnarray}
The exponential of bivectors is useful for defining rotations; a
rotation of vector $a$ by angle $\theta$ on the $\sigma_{12}$ plane
is performed by
\begin{equation}
    a' = \mathrm{e}^{\sigma_{21} \theta/2} a
    \mathrm{e}^{\sigma_{12} \theta/2}= \tilde{R} a R;
\end{equation}
the tilde denotes reversion and reverses the order of all products.
As a check we make $a = \sigma_1$
\begin{eqnarray}
    \mathrm{e}^{-\sigma_{12} \theta/2} \sigma_1
    \mathrm{e}^{\sigma_{12} \theta/2} &=&
    \left(\cos \frac{\theta}{2} - \sigma_{12}
    \sin \frac{\theta}{2}\right) \sigma_1 \nonumber \\
    &&\ast \left(\cos \frac{\theta}{2} + \sigma_{12} \sin
    \frac{\theta}{2}\right)  \\
    &=& \cos \theta \sigma_1 + \sin \theta \sigma_2. \nonumber
\end{eqnarray}
Similarly, if we had made $a = \sigma_2,$ the result would have been
$-\sin \theta \sigma_1 + \cos \theta \sigma_2.$

If we use $B$ to represent a bivector whose plane is normal to
$\sigma_0$ and define its norm by $|B| = (B \tilde{B})^{1/2},$ a
general rotation in 4-space is represented by the rotor
\begin{equation}
    R \equiv e^{-B/2} = \cos\left(\frac{|B|}{2}\right) -  \frac{B}{|B|}
    \sin\left(\frac{|B|}{2}\right).
\end{equation}
The rotation angle is $|B|$ and the rotation plane is defined by
$B.$ A rotor is defined as a unitary even multivector (a multivector
with even grade components only) which squares to unity; we are
particularly interested in rotors with bivector components. It is
more general to define a rotation by a plane (bivector) then by an
axis (vector) because the latter only works in 3D while the former
is applicable in any dimension. When the plane of bivector $B$
contains $\sigma_0$, a similar operation does not produce a rotation
but produces a boost instead. Take for instance $B =  \sigma_{01}
\theta/2$ and define the transformation operator $T = \exp( B)$; a
transformation of the basis vector $\sigma_0$ produces
\begin{eqnarray}
    a' &=& \tilde{T} \sigma_0 T = \mathrm{e}^{-\sigma_{01}\theta/2}
    \sigma_0 \mathrm{e}^{\sigma_{01}\theta/2} \nonumber \\
    &=& \left(\cosh \frac{\theta}{2} - \sigma_{01}
    \sinh \frac{\theta}{2}\right) \sigma_0 \nonumber \\
    &&\ast\left(\cosh \frac{\theta}{2} + \sigma_{01} \sinh
    \frac{\theta}{2}\right) \\
    &=& \cosh \theta \sigma_0 + \sinh \theta \sigma_1. \nonumber
\end{eqnarray}

\section{Reciprocal frame and derivative operators \label{derivatives}}

A reciprocal frame is defined by the condition
\begin{equation}
    \label{eq:recframe}
    g^\alpha \dprod g_\beta = {\delta^\alpha}_\beta.
\end{equation}
Defining $g^{\alpha \beta}$ as the inverse of $g_{\alpha \beta}$,
the matrix product of the two must be the identity matrix, which we
can state as
\begin{equation}
    g^{\alpha \gamma} g_{\beta \gamma} = {\delta^\alpha}_\beta.
\end{equation}
Using the definition (\ref{eq:metrictens}) we have
\begin{equation}
    \left(g^{\alpha \gamma} g_\gamma \right)\dprod g_\beta =
    {\delta^\alpha}_\beta;
\end{equation}
comparing with Eq.\ (\ref{eq:recframe}) we  determine $g^\alpha$
\begin{equation}
    g^\alpha = g^{\alpha \gamma} g_\gamma.
\end{equation}
It would be easy to verify that it is also $g^{\alpha \beta} =
g^\alpha \cdot g^\beta$ and $g_\alpha = g_{\alpha \gamma} g^\gamma$.

In many situations of great interest the frame vectors $g_\alpha$
can be expressed in terms of an orthonormed frame given by Eqs.\
(\ref{eq:basis}). If the frame vectors can be expressed as linear
combination of the orthonormed ones we have
\begin{equation}
    \label{eq:indexframe}
    g_\alpha = {n^\beta}_\alpha \sigma_\beta,
\end{equation}
where ${n^\beta}_\alpha$ is called the \emph{refractive index
tensor} or simply the \emph{refractive index} as said in the main
text. When the refractive index is the identity we have $g_\alpha =
\sigma_\alpha$ for the main or direct frame and $g^0 = -\sigma_0$,
$g^i = \sigma_i$ for the reciprocal frame, so that Eq.\
(\ref{eq:recframe}) is verified.

The first use we will make of the reciprocal frame is for the
definition of two derivative operators. In flat space we define the
vector derivative
\begin{equation}
    \nabla = \sigma^\alpha\partial_\alpha.
\end{equation}
It will be convenient, sometimes, to use vector derivatives in
subspaces of 5D space; these will be denoted by an upper index
before the $\nabla$ and the particular index used determines the
subspace to which the derivative applies; For instance $^m\nabla =
\sigma^m \partial_m = \sigma^1 \partial_1 + \sigma^2 \partial_2 +
\sigma^3 \partial_3.$ In 5-dimensional space it will be useful to
split the vector derivative into its time and 4-dimensional parts
\begin{equation}
    \nabla = -\sigma_0\partial_t + \sigma^i \partial_i
    = -\sigma_0\partial_t
    + \pre{i}\nabla.
\end{equation}
Consistently with the boldface notation for 3-dimensional vectors
$\pre{m}\nabla$ will be denoted by $\bm{\nabla}$. We will use over
arrows, when necessary, to imply that the vector derivative is
applied to a function which is not immediately on its right; for
instance in $\overrightarrow{\nabla}A \overleftarrow{B}$ and in
$\overrightarrow{B}A \overleftarrow{\nabla}$ the derivative operator
is applied to function $B$.

The second derivative operator is called covariant derivative,
sometimes also designated by \emph{Dirac operator}, and it is
defined with recourse to the reciprocal frame $g^\alpha$
\begin{equation}
    \mathrm{D} = g^\alpha\partial_\alpha.
\end{equation}
Taking into account the definition of the reciprocal frame
(\ref{eq:recframe}) we see that the covariant derivative is also a
vector. In cases where there is a refractive index, it will  be
possible to define both derivatives in the same space.

Vector derivatives can also be left or right multiplied with other
vectors or multivectors. For instance, when $\nabla$ is multiplied
by vector $a$ on the right the result comprises scalar and bivector
terms $\nabla a = \nabla \cdot a + \nabla \wprod a.$ The scalar part
can be immediately associated with the divergence and the bivector
part is called the exterior derivative; in the particular case of
Euclidean 3-dimensional space it is possible to define the
$\mathrm{curl}$ of a vector by $\mathrm{curl}(\mathbf{a}) =
-\sigma_{123}\bm{\nabla} \wprod \mathbf{a}.$

We define also second order differential operators, designated
Laplacian and covariant Laplacian respectively, resulting from the
inner product of one derivative operator by itself. The square of a
vector is always a scalar and the vector derivative is no exception,
so the Laplacian is a scalar operator, which consequently acts
separately in each component of a multivector. For $4+1$ flat space
it is
\begin{equation}
    \nabla^2 = -\frac{\partial^2}{\partial t^2} + \pre{i}\nabla^2.
\end{equation}
One sees immediately that a 4-dimensional wave equation is obtained
zeroing the Laplacian of some function
\begin{equation}
    \label{eq:4dwave}
    \nabla^2 \Psi = \left(-\frac{\partial^2}{\partial t^2} +
    \pre{i}\nabla^2\right)\Psi = 0.
\end{equation}
This procedure was used in Ref.\ \cite{Almeida05:4} for the
derivation of special relativity and  extended in Ref.\
\cite{Almeida06:2} to general curved spaces.

\section{Commutation relations \label{commute}}
We examine here the commutation of operators with $\bm{\nabla}$.
First of all we let us expand $\mathbf{x}\wprod \bm{\nabla}$
\begin{eqnarray}
    \mathbf{x}\wprod \bm{\nabla} &=& (x^1 \partial_2 - x^2
    \partial_1)\sigma^{12} + (x^2 \partial_3 - x^3
    \partial_2)\sigma^{23} + \nonumber \\
    && + (x^3 \partial_1 - x^1
    \partial_3)\sigma^{31}.
\end{eqnarray}
The angular momentum operator associated with the plane of 3D
bivector $\mathbf{B}$ is defined as
\begin{equation}
    L_\mathbf{B} = \mathrm{i}\mathbf{B} \dprod (\mathbf{x} \wprod
    \bm{\nabla}).
\end{equation}
For instance, if we are interested in the angular momentum relative
to the $\sigma^3$ direction
\begin{equation}
    L_{\sigma^{12}} = \mathrm{i} \sigma^{12} \dprod (\mathbf{x} \wprod
    \bm{\nabla})
    = \mathrm{i}  (x^2 \partial_1 - x^1 \partial_2).
\end{equation}
The angular momentum operator does not commute with $\bm{\nabla}$;
following Ref.\ \cite{Doran03} we have
\begin{eqnarray}
    [\mathbf{B} \dprod (\mathbf{x} \wprod
    \bm{\nabla}),\bm{\nabla}] &=&
    -\overrightarrow{\bm{\nabla}}\mathbf{B} \dprod
    (\overleftarrow{x}\wprod \bm{\nabla}) \nonumber \\
    &=& \mathbf{B}\times \bm{\nabla}.
\end{eqnarray}
Since $\mathbf{B} \times \bm{\nabla} = (\mathbf{B}\bm{\nabla} -
\bm{\nabla} \mathbf{B})/2$ we can define an operator $\mathbf{B}
\dprod (\mathbf{x} \wprod \bm{\nabla}) - \mathbf{B}/2$ which
commutes with both $\bm{\nabla}$ and $\mathcal{D}$. The conserved
total angular momentum operator is then
\begin{equation}
    J_\mathbf{B} = L_\mathbf{B} - \sfrac{1}{2}\mathrm{i}\mathbf{B}.
\end{equation}
We simplify the notation for the case of bivectors normal to frame
vectors by writing $J_m \equiv J_{\sigma^{no}}$, with $\sigma^m =
\sigma^{123} \sigma^{no}$.

If we ignore second derivatives $(\mathbf{x}\wprod \bm{\nabla})
\bm{\nabla}$ is zero but
\begin{equation}
    \bm{\nabla} ({\mathbf{x}}\wprod \bm{\nabla}) = 2
    \bm{\nabla}.
\end{equation}
For the $\mathcal{K}$ operator we have then
\begin{equation}
    [{\sigma}^{40}(1- \mathbf{x}\wprod \bm{\nabla}), \bm{\nabla}]
    = 2 {\sigma}^{40} \bm{\nabla} -
    {\sigma}^{40} \bm{\nabla} ({\mathbf{x}}\wprod \bm{\nabla}) = 0.
\end{equation}

\end{appendix}

\bibliography{Abrev,aberrations,assistentes}

\end{document}